\documentclass[runningheads]{llncs}
\usepackage{graphicx}
\usepackage{url}
\usepackage{float}
\usepackage{array}
\usepackage{caption}
\usepackage{subcaption}
\usepackage{etoolbox}
\usepackage[table]{xcolor}
\usepackage{enumitem}
\newlist{steps}{enumerate}{1}
\setlist[steps, 1]{label = Step \arabic*:}
\begin{document}
\title{Cryptocurrency wallets: assessment and security
%\thanks{Supported by organization x.}
}
\author{Ehsan~Nowroozi %\inst{1,2}\textsuperscript{(\Letter)}\orcid{0000-0002-5714-8378}
\inst{1} 
%Mauro~Conti \inst{2}\orcid{0000-0002-3612-1934}
, Seyedsadra Seyedshoari\inst{2}, Yassine Mekdad\inst{3}, Erkay Savaş\inst{4}, and Mauro Conti\inst{2}}
\authorrunning{E. Nowroozi et al.}

\institute{Department of Computer Engineering, Bahcesehir University, Istanbul Turkey.\\
\email{ehsan.nowroozi@eng.bau.edu.tr}\\
\and
%Department of Mathematics, University of Padua, 35121, Padua, Italy,
%\email{\{nowroozi,conti\}@math.unipd.it}
%\and
Department of Mathematics, University of Padua, Via Trieste, 63 - Padua, Italy\\
\email{seyedsadra.seyedshoari@studenti.unipd.it, mauro.conti@unipd.it} \\
\and
Cyber-Physical Systems Security Lab, Department of Electrical and Computer Engineering, Florida International University, Miami, FL 33174 USA\\
\email{ymekdad@fiu.edu} \\
\and
Faculty of Engineering and Natural Sciences, Sabanci University, Istanbul, Turkey\\
\email{erkays@sabanciuniv.edu}
}
\maketitle              % typeset the header of the contribu

\begin{abstract}
Digital wallet as a software program or a digital device allows users to conduct various transactions. Hot and cold digital wallets are considered as two types of this wallet. Digital wallets need an online connection fall into the first group, whereas digital wallets can operate without internet connection belong to the second group. Prior to buying a digital wallet, it is important to define for what purpose it will be utilized. The ease with which a mobile phone transaction may be completed in a couple of seconds and the speed with which transactions are executed are reflection of efficiency. One of the most important elements of digital wallets is data organization. Digital wallets are significantly less expensive than classic methods of transaction, which entails various charges and fees. Constantly, demand for their usage is growing due to speed, security, and the ability to conduct transactions between two users without the need of a third party. As the popularity of digital currency wallets grows, the number of security concerns impacting them increases significantly.
The current status of digital wallets on the market, as well as the options for an efficient solution for obtaining and utilizing digital wallets. Finally, the digital wallets’ security and future improvement prospects are discussed in this chapter. 

\end{abstract}

\section {Introduction}
\begin{figure}%
	\centering
	\subfloat[2020]{{\includegraphics[width=.53\textwidth,height=4.3cm]{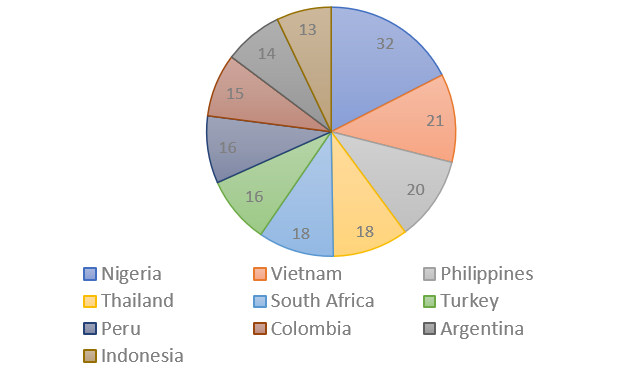} }}%
	%\qquad
	\subfloat[2021]{{\includegraphics[width=.53\textwidth,height=4.3cm]{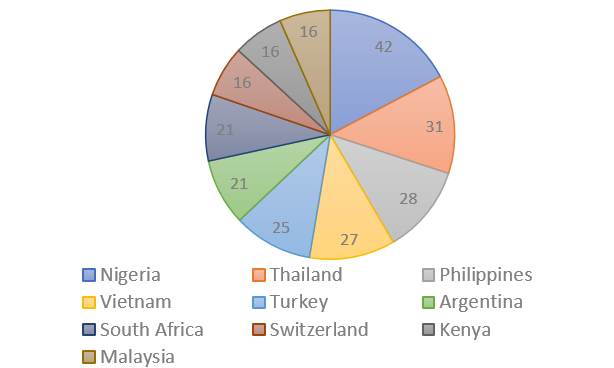} }}%
	\caption{The percentage of people owning or using cryptocurrencies in different countries. (a) Shows the percentage in 2020, and (b) Shows the percentage in 2021 \cite{Countries}.}%
	\label{fig:countries graphs}%
\end{figure}

\begin{figure}[h!]
  \centering
 % \captionsetup{justification=centering,margin=0.4cm}
  \includegraphics[width=20pc]{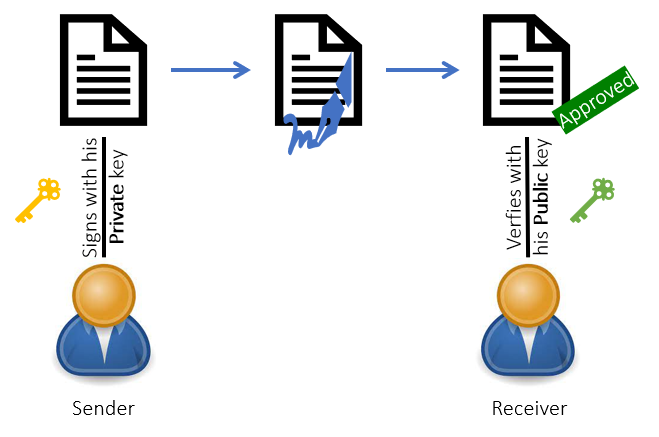}\\
  \caption{Blockchain public/private key cryptography.}
  \label{fig:Keyss}
\end{figure}

Since bitcoin's introduction, the number of cryptocurrencies has increased to thousands, with a market valuation of over \$1.72 trillion as of March 2022 \cite{Marketcap}. Meanwhile, the percentage of people owning or using cryptocurrencies is growing significantly in many countries as shown in Figure~\ref{fig:countries graphs}. A digital wallet as a software application for cryptocurrencies, keeps private/public keys and properly operates on various blockchains, allowing users to transfer currencies to each other and monitor their currency balance eliminating need for a physical wallet \cite{Singh}. Blockchain-based cryptocurrencies are built on the blockchain concept, which is a decentralized open database with entries that may be verified but not modified \cite{Daojing}. Various currencies could be stored, sent, and received using a digital wallet. Within the wallet, cryptocurrencies are not kept like real money. The blockchain captures and archives every transaction \cite{CryptoWallets}. A wallet transaction involves sending currency between two addresses. Private key of the sender and public key of the receiver is required for a transaction to take place \cite{Bitcoin}. Any quantity of coins owned by the sender can be transferred to public key (or address) of the receiver. To verify that the transaction was started and performed by the sender, it digitally signs the transactions using its private key as shown in the Figure ~\ref{fig:Keyss}. The mainnet includes both the sender and the recipient, which is where transactions take place. There is a separate network called testnet that is utilized for testing, however, testnet coins have no actual worth. Users cannot transmit cryptocurrencies between the mainnet and the testnet since they are two independent networks.
 In principle, bitcoin wallet applications establish new addresses, securely keep private keys, and assist in the automating of transactions. Several wallets only accept one type of cryptocurrency (for instance, Bitcoin), whereas others such as Exodus and Jaxx support a wide variety of cryptocurrencies.\\
Cryptography is a field of study that deals with above-mentioned keys. Cryptocurrency consists of two security approaches, symmetric and asymmetric. First one has a secret key whereas the second model contains public and private keys. Encryption and decryption in symmetric mode are simply done by utilizing traditional symmetric encryption techniques like Data Encryption Standard (DES), where the same key is used for encryption and decryption \cite{Aydar}.\\

Asymmetric encryption often used in cryptocurrency exchanges, is an encryption technique that employs two keys (public and private keys) paired with distinct encryption and decryption methods. Although the sender must have a duplicate of the recipient's public key in order to transmit a coin, it should be assumed that the adversary has the exact copy. In this case, the sender encrypts the message with the proper encryption mechanism, which the recipient of the message may decode using its private key. The purpose of the asymmetric approach is to prevent an attacker from utilizing the public key to decrypt an encrypted communication \cite{CryptoWallets} \cite{Kriptologija}. 
\\
\\
\subsection{Crypto wallets' Categories}
Digital wallets based on their features like online/offline working mode can be divided into two categories: hot and cold wallets. Their main distinction is that a connection to the internet is required for hot wallets, whilst cold wallets do not. Users of a hot wallet typically use it to do online purchases and for that reason, users should not allocate a large amount of money, but a cold wallet functions similarly to a bank vault for storing various digital assets.
It's advisable to have both wallets, mostly for security purposes \cite{CryptoWallets}. 
\\
There are some different types of hot and cold wallets. Desktop wallets, hardware wallets, mobile wallets, online wallets, and paper wallets are all available. Hot wallets include multi-signature, desktop, mobile, and online wallets, whereas cold wallets include paper and hardware wallets. Each cryptocurrency wallet has its own level of safety and privacy to ensure that the private key is kept securely.\\
\\
Each kind of digital wallets and their advantages and disadvantages have been described as follows \cite{CryptoWallets}:\\

\textbf{Mobile wallets:} Mobile wallets are more efficient and simple than using other kinds of crypto wallets since they can be accessed from anywhere via an internet connection. Despite the fact that new mobile wallets take advantage of the security mechanisms of smartphones like ARM TrustZone to protect users \cite{TrustZone}, it is susceptible to viruses and hacking. This method allows users to use the TOR network for increased security. TOR is a common anonymous communication network with a low rate of delay that allows users to connect online resources without disclosing their network id \cite{TOR}. Another fantastic feature is the ability to scan QR codes. Mobile phones, on the other hand, could be considered as unsafe equipment. As a result, if the phone gets compromised, the user's crypto tokens may be lost. They are also vulnerable to viruses, malware, and key loggers.\\

\textbf{Online wallets:} This form of wallet may be accessed through any web browser without the need to download or install any software. Since these wallets are susceptible, it is not suggested that users keep a high number of cryptocurrencies in them. Cryptocurrency transfers are conducted in a timely manner. It is advised to hold a little number of cryptocurrencies in these wallets. \\

Several numbers of this digital wallet are capable of holding multiple cryptocurrencies as well as transferring funds amongst them. It allows customers to use the TOR network for more confidentiality and privacy. However, a  third entity or centralized administration has complete control over the digital wallet. It is suggested to utilize a personal computer (PC) with a necessary security application pre-installed in order to use an online wallet. Users are vulnerable to different online scams due to a lack of awareness of information technology (IT). \\

\textbf{Desktop wallets:} Desktop wallets are assumed to be more secure than mobile and online wallets, however, this might vary depending on the level of protection for online crypto wallets' security. Although a desktop wallet can create addresses for receiving cryptocurrency offline, it requires the use of the internet to send them out. Transaction logs will not be refreshed if there is no internet connection \cite{Taylor}. Although these wallets are simple to use and keep private keys on the user's device, a machine connected to the internet becomes insecure and demands additional protection and security. Furthermore, frequent backups are required because the system may fail at any time, resulting in the loss of all data. Otherwise, the user needs to export the related private key or seed phrase. As a result, users will be able to access digital content on several devices \cite{Review}.\\

\textbf{Multisignature wallets:} Based on the level of protection, two or three private keys are required to access money after conducting a transfer using a multi-signature wallet. This method is beneficial to businesses because it allows them to delegate responsibility to many staff, who must all provide their own private key in order to access assets. BitGo is an instance of a multi-signature wallet, where the users store the first key, a trusted third party stores the second key, and the firm itself stores the third key. Transactions might be slow due to the number of signatures required. Multisignature relies on the signing of the transaction by additional devices or a third party.\\

\textbf{Paper wallets:} This is one of the most secure wallets available. They fall in the category of cold crypto wallets. A paper wallet, as the name implies, is a printed sheet of paper containing both private and public keys. \\

A QR code is printed on the paper, which indicates the keys of the user and may be used for almost any kind of transaction. The user's principal attention should be retaining that paper safe, as the result, this wallet is the safest. They are kept in the physical wallet or pocket of the users without requiring a connection to the computer; however, the transaction takes longer to be completed. \\

\begin{figure}[H]
  \centering
 % \captionsetup{justification=centering,margin=0.4cm}
  \includegraphics[width=7pc]{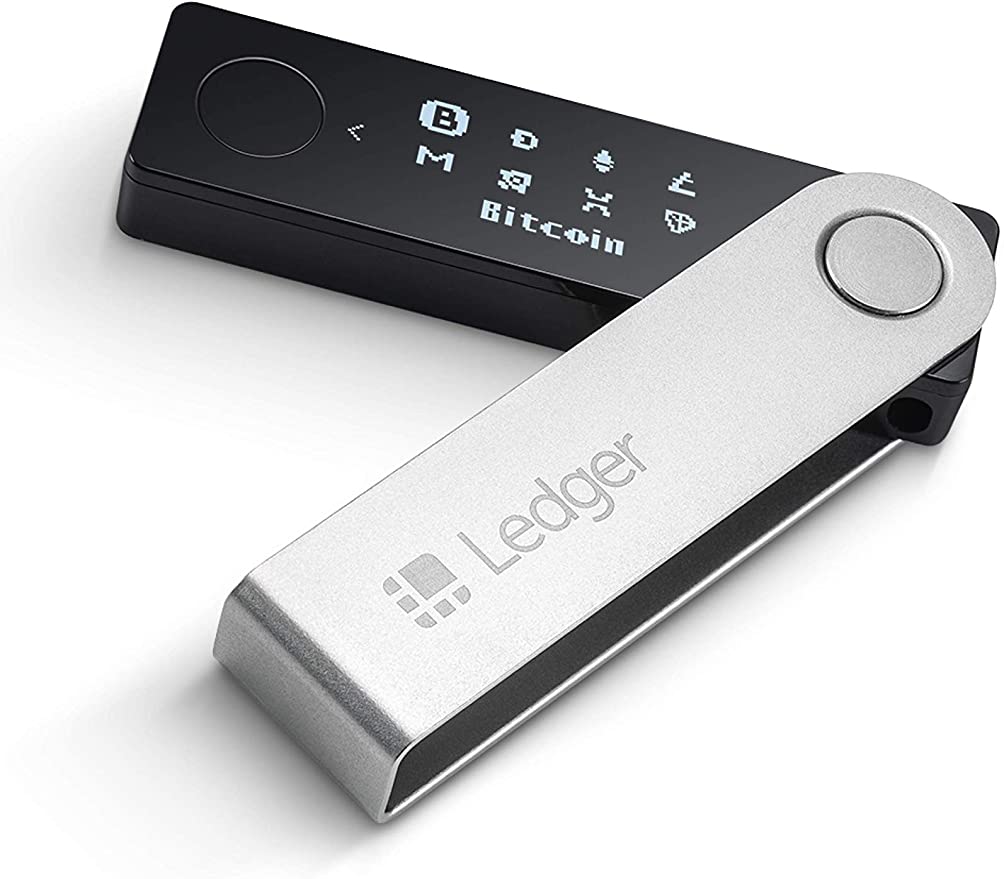}\\
  \caption{Ledger Nano X, portable hardware wallet \cite{HWWallets}}
  \label{fig:NanoX}
\end{figure}

\textbf{Hardware wallets:} These wallets are specialized devices of cryptography for generating, storing private keys, and authenticating transactions \cite{SecApproach}. In most instances, they are safer wallets because transaction signing occurs on the hardware wallet, and the private key does not leave the safe hardware wallet system, it prohibits malware from stealing digital wallets \cite{Horus}. 
A hardware wallet is commonly a USB flash memory (Figure ~\ref{fig:NanoX}) with software installed and ready to use. Some of these devices contain a screen, allowing the user to conduct a transaction without the need of a computer. This kind of wallet provides the user with additional control over their cryptocurrency and is an appropriate option for the long-term storage of crypto assets. The majority of secured USB wallets have a screen. They are safer than all other sorts of digital wallets. They are, however, quite tough to get and are not suggested for novices. \\
A comparison of different wallets is provided in Table ~\ref{Tab:Comparison}.
\\

\begin{table}[h!]
    \begin{center}
    \caption{Comparison of different categories of cryptowallets}
    \AtBeginEnvironment{tabular}{\small}
    \newcolumntype{M}[1]{>{\raggedright\let\newline\\\arraybackslash\hspace{0pt}}m{#1}}
    \newcolumntype{N}[1]{>{\raggedright\let\newline\\\arraybackslash\hspace{0pt}}m{#1}}
    \newcolumntype{O}[1]{>{\raggedright\let\newline\\\arraybackslash\hspace{0pt}}m{#1}}
    \newcolumntype{P}[1]{>{\raggedright\let\newline\\\arraybackslash\hspace{0pt}}m{#1}}
    \label{Tab:Comparison}
    \begin{tabular}{| p{3cm} | p{3cm} | p{3cm} |}
    
    \hline
    Wallet & Advantages & Disadvantages \\
    
    \hline
    Mobile wallets&
    - Efficiency and simplicity of use \newline -~Supporting TOR network\newline - Using QR code& 
    - Loss of crypto tokens due to compromising the phone\newline - Prone to key logger, viruses, and malware\\
    \hline
    
    \hline
    Online wallets&
    - Fast transactions\newline - Supporting TOR network\newline - Supporting multiple cryptocurrencies and transactions between them & - Fully controlled by central authorities or third party\newline - Demanding a personal computer and installing specific application\\
    \hline
    
    \hline
    Desktop wallets&
    - Simplicity of use\newline - Storing private key on user's system & 
    - Susceptible and requiring more security\newline - Regular backup required\\
    \hline
    
    \hline
    Multisignature wallets &
    - Dedicating responsibility to employees of a company & 
    - Slow transactions\\
    \hline
    
    \hline
    Paper wallets&
    - Kept in user's packet or physical wallet& 
    - Slow transactions\\
    \hline
    
    \hline
    Hardware wallets &
    - LCD screen on USB wallets\newline - Safer than others& 
    - Hard to purchase\newline - Not suggested for beginners\\
    \hline

    \end{tabular}
    \end{center}
\end{table}

Users who intend to trade in several currencies may consider multi-currency wallets. Although Bitcoin is the most well-known currency, there is a large number of other cryptocurrencies on the market, each with its own infrastructure network \cite{Laptop}. \\

\subsection{Available Digital Wallets}
It is important to keep in mind that cryptocurrency is outlawed or restricted in certain states and countries prior to deciding on a digital currency, while its usage and exchange are permitted in others. It is advisable to select a multi-currency wallet that supports several cryptocurrencies \cite{Review}. It is possible to lose money by selecting the incorrect wallet for a certain digital currency. Users should spend some time learning about the various types of cryptocurrency wallets and their functionality.
In this section some of the most common wallets are listed as follows: Exodus (online wallet), Coinpayments (online wallet), Ledger Nano S (hardware wallet), Jaxx (mobile wallet), and Ledger Blue (hardware wallet) \cite{CryptoWallets}.\\

\begin{figure}[h!]
  \centering
 % \captionsetup{justification=centering,margin=0.4cm}
  \includegraphics[width=20pc]{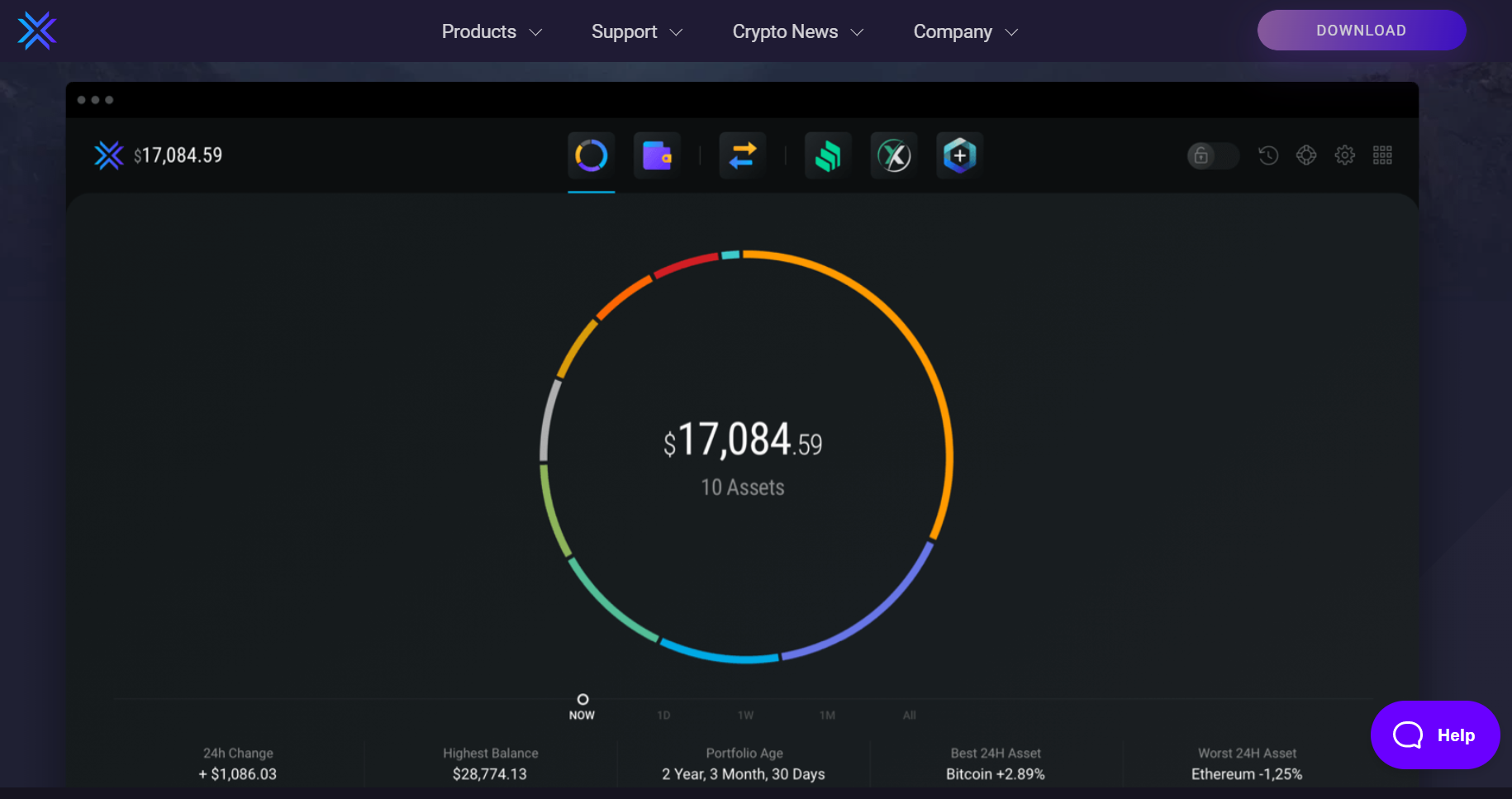}\\
  \caption{Exodus platform}
  \label{fig:Exodus}
\end{figure}

\textbf{Exodus} is a web-based electronic wallet with a user-friendly interface shown in Figure ~\ref{fig:Exodus}, a stylish design, and a reporting mechanism. When compared to other online wallets, Exodus provides comparable functionality, with some being better than others. In this type of wallets, registration is free of charge so anyone may submit the form and become the owner of a crypto-wallet of this type. The cryptocurrency swap, where users can trade several crypto assets without incurring extra charges, is one of its best features %\cite{mobile_wallets}. 
It is a fantastic place for inexperienced traders. Although it is an online wallet, it is also an offline wallet since the data is kept on the computer of the user when the wallet is generated \cite{CryptoWallets}.\\

\begin{figure}[h!]
  \centering
 % \captionsetup{justification=centering,margin=0.4cm}
  \includegraphics[width=17pc]{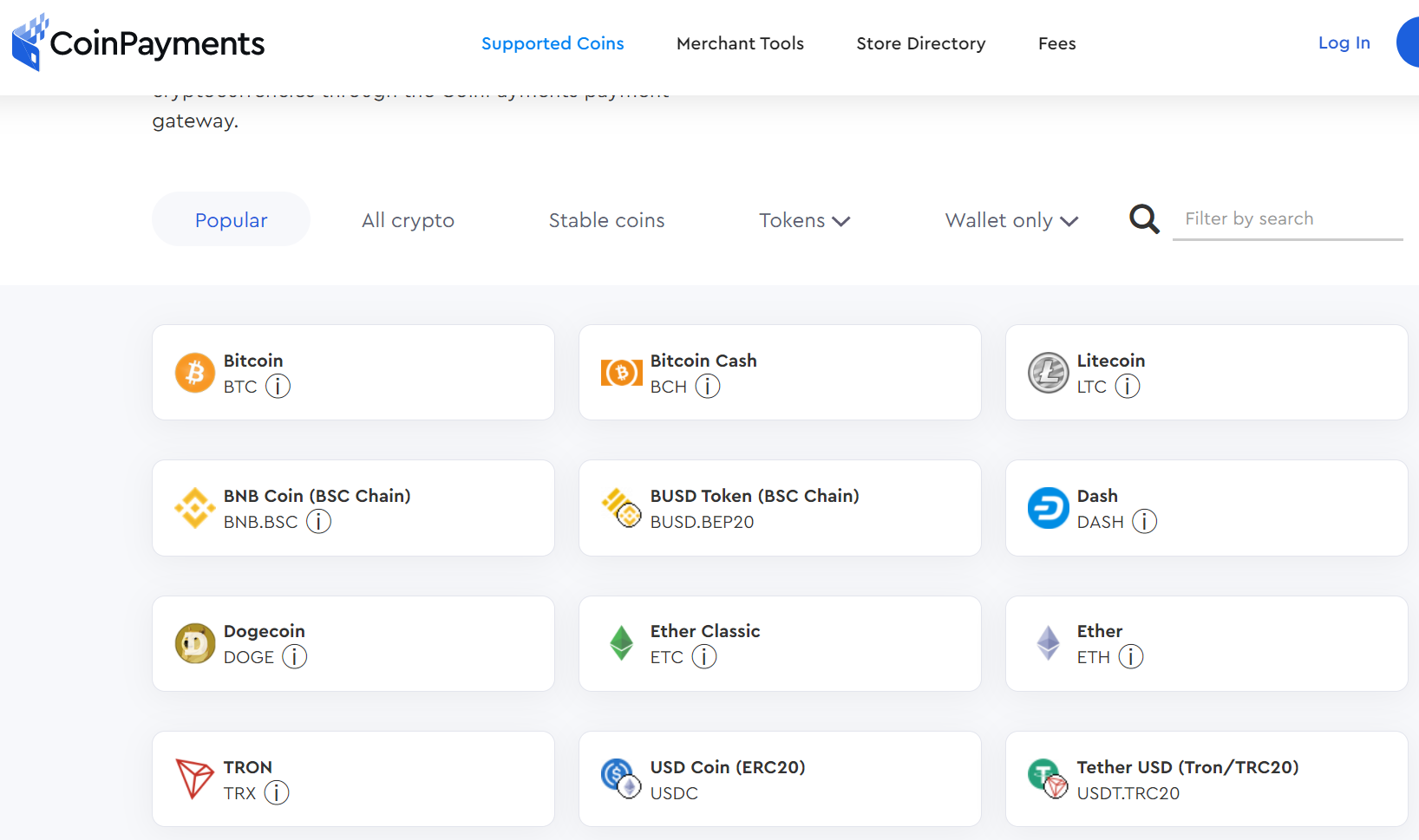}\\
  \caption{Coinpayments platform}
  \label{fig:Coinpayments}
\end{figure}

\textbf{Coinpayments} is a digital wallet that can be accessed online. They become popular after proving that their wallet could hold at least 300 various digital currencies as illustrated in Figure~\ref{fig:Coinpayments}. They only get paid when a user finalizes a transaction using their wallet. Because this wallet accepts multiple currencies and is accepted by so many online retailers, it is feasible to shop online using this wallet. The BitGo services have been integrated into this wallet to provide a higher level of security and transaction speed. Moreover, a safety function is added to keep  the money of the users safe from criminals. This wallet allows users to store several currencies in the same place. In addition, a lot of online retailers utilize it for online shopping.\\

\textbf{Ledger Nano S} is a digitized USB wallet for cryptocurrencies introduced in 2016. Due to the fact that hardware wallets are significantly more expensive than other digital wallets, but they are a cost-effective investiture with a variety of capabilities such as enabling the users to securely trade and monitor digital assets as well as supporting more than 1,100 cryptocurrencies and tokens \cite{Invest}. \\

\begin{figure}[h!]
  \centering
 % \captionsetup{justification=centering,margin=0.4cm}
  \includegraphics[width=10pc]{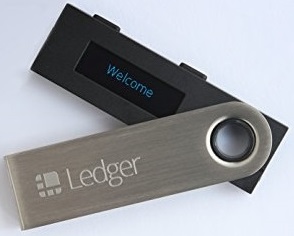}\\
  \caption{Ledger Nano S device}
  \label{fig:LedgerS}
\end{figure}

The private key's backup and security are given special consideration. This gadget can be started without the need for a computer. It includes a little LCD screen on the front of the USB so that the users can use it easily as shown in Figure~\ref{fig:LedgerS}. It makes it possible to move money from one account to another as well as exchange cryptocurrencies. There are two sizes of Ledger Nano S, the larger device is 98 mm, while the smaller device is 60 mm. This wallet can hold a variety of famous cryptocurrencies. The user may keep an eye on current transactions and utilize the button to double-chseck them. Several security features are available, as well as the possibility to lock the wallet using a password. Regardless of how little the gadget is, it can be conveniently utilized by users. \\

\begin{figure}[h!]
  \centering
 % \captionsetup{justification=centering,margin=0.4cm}
  \includegraphics[width=14pc]{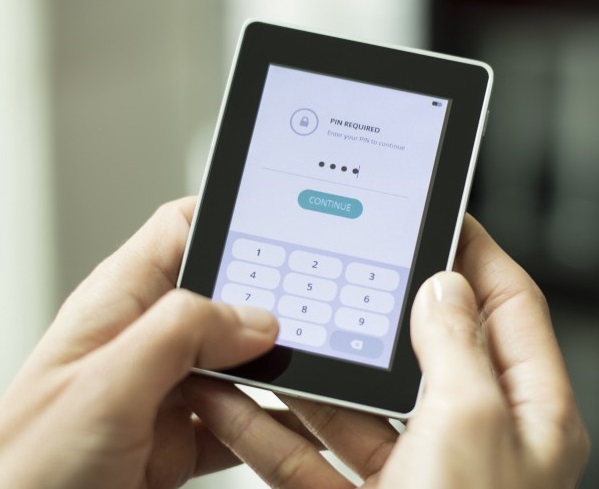}\\
  \caption{Ledger Blue device \cite{LedgerBlue}}
  \label{fig:LedgerBlue}
\end{figure}

\textbf{Ledger Blue} is also a hardware wallet designed by the same company. It outperforms the Ledger Nano S and adds plenty of additional features which could be seen in Figure~\ref{fig:LedgerBlue}. This wallet is one of the most costly wallets on the market due to these qualities. To prohibit external access, the users can specify a code with 4 to 6 digits. The Ledger blue wallet uses dual-chip technology and includes built-in software for digital currency safety. It is completely immune to harmful malware which means attackers cannot hack it.\\

\begin{figure}[h!]
  \centering
 % \captionsetup{justification=centering,margin=0.4cm}
  \includegraphics[width=10pc]{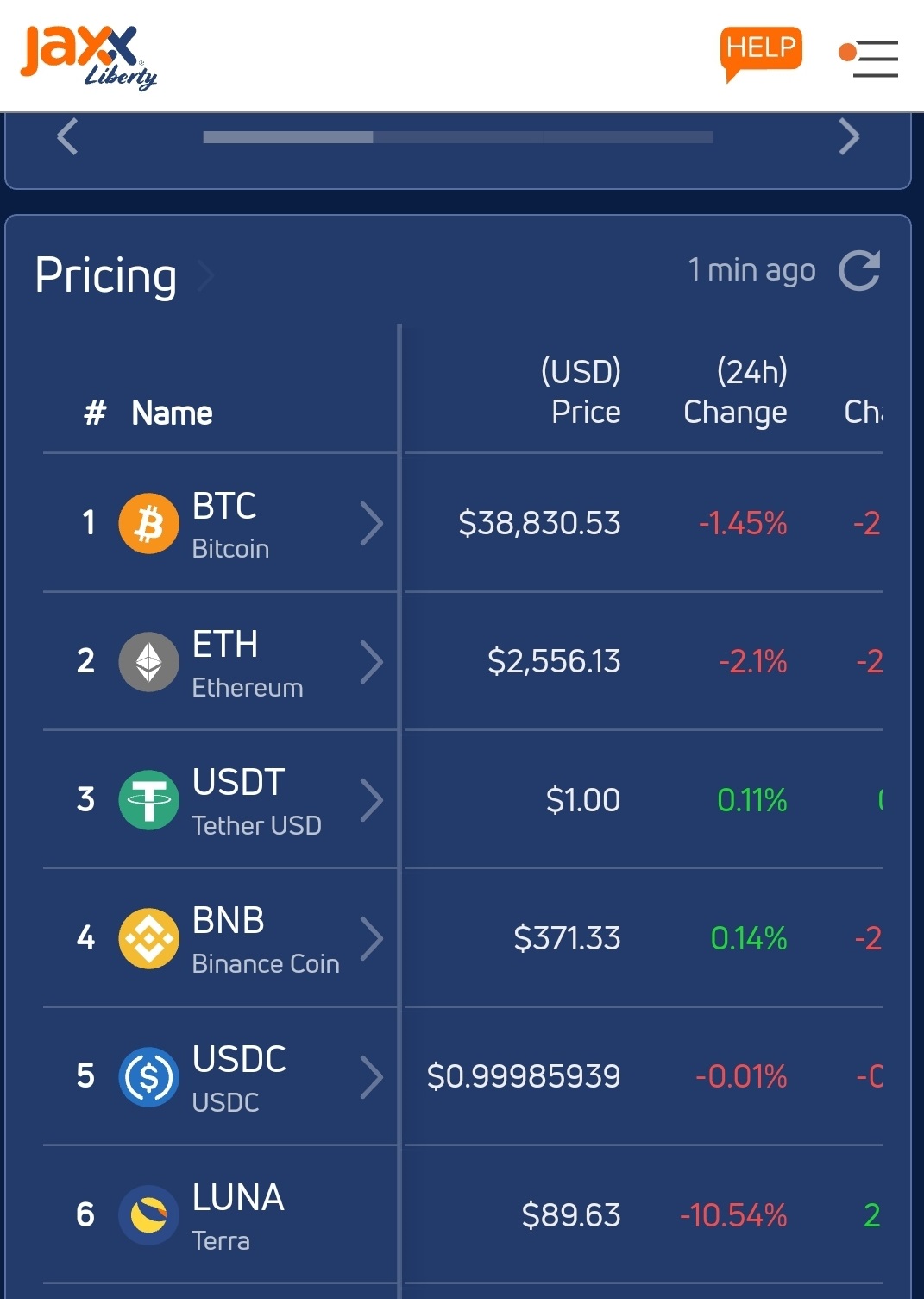}\\
  \caption{Jaxx Liberty platform}
  \label{fig:Jaxx}
\end{figure}

\begin{figure}[h!]
  \centering
 % \captionsetup{justification=centering,margin=0.4cm}
  \includegraphics[width=17pc]{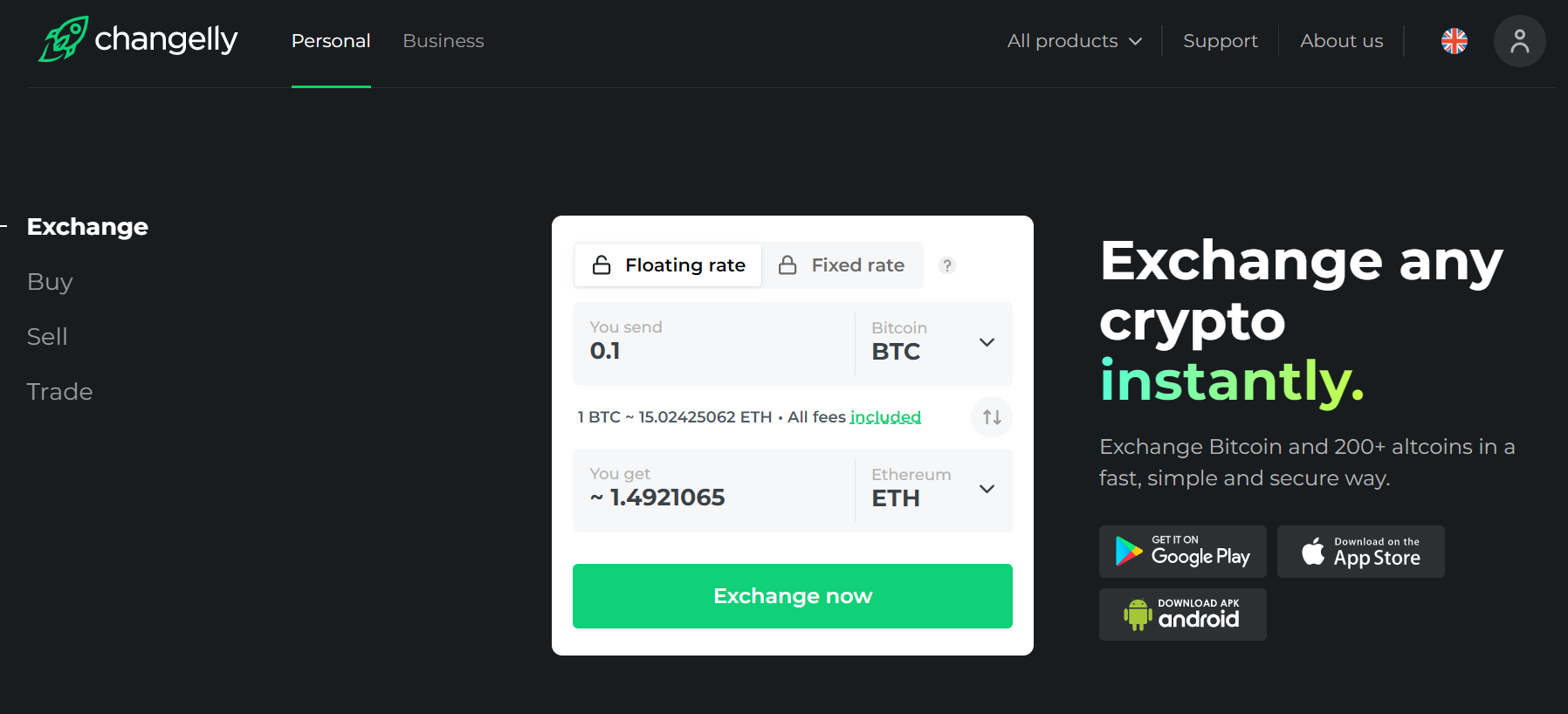}\\
  \caption{Changelly exchange platform}
  \label{fig:Changelly}
\end{figure}

\textbf{Jaxx Liberty} is one of the mobile and web digital wallets (illustrated on Figure~\ref{fig:Jaxx}) that may also be referred to as a desktop wallet since it operates on both Windows and mobile platforms that allows the user to trade digital assets using third-party services like Changelly as shown on Figure~\ref{fig:Changelly}. For all digital assets, Jaxx was intended to keep them safe from cybercriminals. New mobile wallets offer a variety of security measures if a user's phone is lost. If this is the case, they'll let users swap accounts. Jaxx is compatible with all main operating systems such as Android, IOS, Windows, Linux, and Mac OS. Jaxx enterprise is not able to view the user's digital currency since a private key is produced and saved on the computer of the user. In most cases, making a transaction with an online wallet requires a number of procedures. The Jaxx concept is focused on the Nada privacy model. Nada is responsible for protecting confidentiality and privacy.\\
Main features of discussed wallets are summarized in Table ~\ref{Tab:Features}.

\begin{table}[h!]
    \begin{center}
    \caption{Main features of some of most common wallets}
    \AtBeginEnvironment{tabular}{\small}
    \newcolumntype{M}[1]{>{\raggedright\let\newline\\\arraybackslash\hspace{0pt}}m{#1}}
    \newcolumntype{N}[1]{>{\raggedright\let\newline\\\arraybackslash\hspace{0pt}}m{#1}}
    \newcolumntype{O}[1]{>{\raggedright\let\newline\\\arraybackslash\hspace{0pt}}m{#1}}
    \newcolumntype{P}[1]{>{\raggedright\let\newline\\\arraybackslash\hspace{0pt}}m{#1}}
    \label{Tab:Features}
    \begin{tabular}{| p{2.04cm} | p{9cm} |}
    
    \hline
    Wallet & Main features\\
    
    \hline
    Exodus&
    \textbf{Safe:} Information are saved on user's system while being created\newline \textbf{Multi-currency:} Supports diverse currencies in the same wallet\newline \textbf{Free registration:} Possessing this wallet by simply filling out a form\\
    \hline
    
    \hline
    Coinpayments&
    \textbf{Safe:} Money of user is protected against stealing\newline \textbf{Multi-currency:} Supports diverse currencies in the same wallet\newline \textbf{Integrated with BitGo services:} Increases speed and security of transactions\newline \textbf{Common:} Used by thousands of online shops\\
    \hline
    
    \hline
    Ledger Nano S&
    \textbf{User-friendly:} Could be used comfortably\newline \textbf{Multi-currency:} Supports diverse currencies in the same wallet\newline \textbf{Small screen:} Monitoring current transactions and confirming them by a button\newline \textbf{Backup and recovery:} Fast restoring process if digital money is lost\newline \textbf{Safe:} Provides many security features like a password to lock the wallet\\
    \hline
    
    \hline
    Ledger Blue&
    \textbf{Pin code:} Restricts external access using a password with 4-6 characters\newline \textbf{Resistant to malicious software:} Cannot be violated by malwares\newline \textbf{Safe:} Benefits from dual chip design and includes a firmware for security \\
    \hline
    
    \hline
    Jaxx&
    \textbf{Acceptable:} Could be implemented on any OS\newline \textbf{Easily operated:} Does not need a lot of steps to execute a transaction
    \newline \textbf{Full control:} Private key is stored and accessed only on user's computer\newline \\
    \hline

    \end{tabular}
    \end{center}
\end{table}

\section{Overview of Digital Wallets' Security}
The crypto wallets' security goals, including availability, integrity, and confidentiality, are compatible with most security standards. Adversary makes use of vulnerabilities in wallet libraries to create a distinctive impression of the wallet finger that is linked to the user's identity for further monitoring. Although, Android and IOS provides a variety of tools for programmers and customers, some of these features can be abused by hackers to violate the security of the framework of cryptocurrencies that runs on the platform \cite{Varghese}. 
\\
There are many data transmission functions in the bitcoin cryptocurrency, the most prominent crypto money. These features may pose a security risk, but Bitcoin has a very strong security system, implying that they should be used correctly. The security of the platform should be a priority while investing in online platforms. When purchasing a wallet for this digital money, two-factor verifications are suggested which is a process of authentication that requires two assets of the users including what they know such as login credentials, and what they have like a mobile phone to receive a one-time password (OTP) \cite{Two-factor}.
In comparison to a physical wallet, a smart approach to storing money in the wallet is possible. This implies that a little quantity of currency should be kept in the digital crypto wallet for daily utilization. \\
\\
Backup wallets can help eliminate issues such as information theft or errors in the computer; however, this condition can only be met if the data has been encoded. The security of data saved on the network is not completely guaranteed. Malicious software is able to infect any machine connected to the internet. Encryption of the data is essential to eliminate any risk of being compromised, which is a vital safety measure. Data should be kept in a multitude of places like a backup wallet. It's not just about cloud storage when it comes to various places, but also regarding physical devices like CDs, external hard disks, USBs, and so on. Daily or frequent backups guarantee that the data is constantly updated. When it comes to digital wallets, encryption plays a critical role. Thus, encrypting the user's cryptocurrency wallet is quite effective method to protect the money saved within that wallet. When someone attempts to enter the digital wallet, a password is created. The password should not be forgotten or lost, as this would result in the loss of funds. The distinction between actual money and cryptocurrency is that if a user's password is lost, may obtain a new one. The user has complete responsibility in cryptocurrency and blockchain. It's critical to combine characters, numbers, and letters to set up a secure password.\\
\subsection{Cryptocurrency Wallet's Backup}
A backup wallet is simply another term for transferring money to another location or producing a replica \cite{CryptoWallets}. There should be two wallets during the backup process: primary wallet and backup wallet which is working offline \cite{Multilayered}. To keep safe cryptocurrencies, encrypted local backups of the funds can be saved in a hardware wallet that is not connected to the internet (backup wallet). Existing hardware wallets' backup and recovery procedure is a significant problem since most of them employ a word list (mnemonics) to create a duplication of the private keys and restore them while required. These words must be written on paper and kept secure by the user \cite{SecApproach}. This approach pushes the problem outside the wallet by converting the private key's seed from digitized to physical type. Rezaeighaleh et al. \cite{SecApproach} proposed a novel framework for a backup and recovery process implementing Elliptic-Curve Diffie-Hellman (ECDH) algorithm, which could be easily used by users since they no longer need to write the word list and save it.\\ 
\subsection{Cryptocurrency Wallet's Encryption}
Encrypting confidential information and digital money has always been a robust and reliable method of protection. In cryptocurrency, hashing is a way of transforming huge amounts of data into little digits. It is used in the Bitcoin network for encoding the wallet's address, encoding transactions between two wallets, and confirming the balance in a specific wallet. The Bitcoin network employs a safe hash algorithm such as SHA-256. One of the most important features of this technique is that modification of one bit of incoming data will totally alter the output. This is related to the Avalanche effect, which reflects the behavior of traditional mathematical algorithms such as Data Encryption Standard (DES) and Advance Encryption Standard (AES), in which a little variation in the input causes the entire hash value to alter considerably \cite{Comparative}. A slight modification in the plaintext may result in a large shift in the ciphertext when using symmetric ciphers. Otherwise, an error will appear while decrypting the encrypted text \cite{Palette}.
Public key cryptography is a technique of proving the identity of a person by using a pair of cryptographic keys (a private key and a public key). A digital signature is created by combining both keys. The blockchain wallet, which connects with the blockchain network, stores these private keys, public keys, and blockchain addresses as well as keeps track of the coins that could be transmitted via digital signature \cite{Varghese}.\\

Since possession of the private key entails complete control over the related cryptocurrency account, managing the private key is critical for security. Before saving in the wallet, the private key must be encoded, and when used, must be decoded into plaintext. The plaintext of the private key in Ethereum, for instance, is a 256-bit binary integer that is usually displayed encoded as a hexadecimal number. Before being saved in the wallet, the private key must be encrypted and then decrypted if needed. \\

\subsection{Cold Wallets as Another Solution}
Cold wallets are another solution for storing and protecting data. These are hardware wallets that do not demand an online connection and use a USB stick to transfer transactions and keys \cite{SecApproach}. Two computers share some parts of the same digital wallet while signing transactions offline. Only the first computer must be disconnected from all networks, as it is the only one with an entire digital wallet and permission for signing transactions. The other computer is connected to the internet and holds the digital wallet, which can only be used to observe and execute unsigned transactions. Only a few steps are required to complete the transaction:
\begin{itemize}
\item \textbf{Step 1}: A new transaction should be created on the computer with a connection to the internet and save to a USB device.\\

\item \textbf{Step 2}: Transaction must be signed with the computer which does not have an internet connection.\\

\item \textbf{Step 3}: Signed transaction should be sent with the computer which is connected to the network.
\end{itemize}

\subsection{Cryptocurrency Wallets and QR Code}
Ghaffar Khan et al. \cite{Khan} employed QR codes for cross-verification across hot and cold wallets to keep digital currencies. Cold wallets are safer against cyber-attacks due to their offline nature; This approach is like an additional protection layer of bitcoin transactions \cite{Dikshit}. All cryptocurrency investors should understand the differences between hot and cold wallets in order to ensure safe and secure digital money transactions. Online wallets can send the funds and distribute them in a network only after confirming the private key of the cold wallet and scanning the QR code. 

The version of the digital wallet application must be upgraded on a regular basis since every time the program is updated, the users receive vital security upgrades. Updates may provide new capabilities for crypto-wallets, as well as the prevention of a number of issues with different degrees of intensity. Numerous signatures can be used in crypto wallets, requiring several confirmations before a transaction can be funded. This form of security may be employed in larger businesses like banks with staff who have access to government coffers. The multi-signature feature is also available in some web wallets such as BitGo, and Coinbase \cite{Laptop}. \\

\section{Cryptocurrency Wallets' Security Objectives}
Cryptocurrency wallets have security goals that are similar to those of other security structures, including availability, integrity, and confidentiality \cite{Daojing}. \\

\textbf{Availability:} The purpose of availability is to guarantee that the legal use of data is not inhibited, which means that the information must be usable and available while demanded by a valid authority. It's critical for wallet applications to make sure that keys can be produced, saved, and retrieved appropriately. In addition, transactions should be properly signed, transmitted, and accessed in response to user queries \cite{Daojing}. \\
The wallets can become unavailable if any failure, overload, or attack occurs. Important features of availability are fail-safe, reliability, scalability, fault-tolerance, up-time, and recoverability. The system could be called fail-safe if the attack or failure has the least impact such as data loss. Reliability is known as the probability of operating as expected if no outside source attempts to interrupt the system. A scalable system allows for increasing the number of available resources without modifying the system architecture. A system can be assumed as a fault-tolerance system if it is able to continue operating properly even with a decreased level of functionality. Up-time is referred to the period of time that the system is actively working and accessible to users. Finally,the term "recoverability" refers to the ability of a system to recover its data in an acceptable time frame in the event of a breakdown \cite{Shojae}.\\

\textbf{Integrity:} Integrity refers to the ability to prohibit illegitimate entities from altering data in order to ensure its completeness and correctness. When it comes to blockchain wallets, ensuring the integrity of the private key is critical. The user will lose his/her account's control if the private key kept in the wallet gets modified or deleted in an illegal way, resulting in the loss of the account's assets. Blockchain has employed cryptographic methods like hashes and signatures to verify that transaction data has not been changed before being transmitted to the blockchain. The integrity feature, on the other hand, is critical for a recently launched transaction. Even if the transaction's data has been altered before being signed by user with the private key, the transaction will be validated by the blockchain system since it carries the signature of the legal owner. It's also possible to tamper with historical transactions once they have been retrieved from the blockchain system. \\

\textbf{Confidentiality:} The goal of confidentiality is to keep sensitive information safe from unwanted access. A digital currency account's private key grants complete control over the account and any digital assets held within it. As a result, the wallet's primary security feature is to guarantee that the private key is not accessible in an illegal manner. Because all the information is publicly available on the blockchain, transaction information is not assumed to be confidential.

\section{Cryptocurrency Wallets' Adversary Model}
Various sorts of digital money wallets have different adversary models like the application-oriented adversary model and physical access adversary model \cite{Multilayered}. In this section, the adversary model for cryptocurrency wallets based on software has been discussed. The purpose of the adversary is to compromise the availability, integrity, or confidentiality of the wallet's data. This involves tampering with earlier transactions, preventing the initiation of new transactions, accessing the private key, manipulating newly launched transactions, refusing transaction information queries, etc \cite{Daojing}. \\

The attacker lacks private information specific to a target wallet's owner, like the list of wallet transaction passwords or the user's account's private key. The attacker, however, has the potential to install and execute any program that is installed on the same system as the wallet operates. All the permissions requested by the installed program have been granted. Any option on the device where the wallet operates can be changed by the attacker. The attacker can also execute any program on the user's other devices that utilizes the wallet. The wallet's communication can be listened to and modified by the adversary, even if they don't have access to the encrypted traffic's key. The servers connected to the wallets can be attacked by the adversary, but the blockchain network cannot be controlled by attackers.\\

The adversary approach described above is realistic since the users might be persuaded to install a new program and then provide it with all the necessary permissions. The program can imitate the appearance of a standard program. Furthermore, tactics like accessibility services, USB debugging, as well as other smartphone functions might provide attackers with extra possibilities to exploit \cite{Daojing}. 

\section{Vulnerabilities in Cryptocurrency Wallets}
Transaction management and private key management are two of the most important functionalities of cryptocurrency wallets. Transaction management comprises sending and gathering tokens, as well as querying balances and transactions, while key management covers a private key's creating, saving, importing, and exporting; however, if these capabilities are used incorrectly, attack points may be introduced into the attack surfaces. Furthermore, because an operating system (OS) hosts the digital wallet,  an attacker might be able to exploit the OS's properties, arising a danger to the digital wallet's security \cite{Daojing}. \\

The attack surface from the perspective of the cryptocurrency wallet and its underlying operating system has been discussed in the following.
\subsection{Cryptocurrency Wallets' Attack Surface}

\textbf{Transaction Management:} While a user intends to withdraw money from an account, the wallet creates a transaction and signs it using the user's private key. Then, it sends the signed transaction to the blockchain system for confirmation in order to accomplish the operation of the transaction. When a user has to perform a collection process, must present the payer address of their account, which might contain the currency and amount. \\Users can access the related account balances and account transaction logs using transaction records of the wallet application and balance inquiry services. This approach may need a connection to the server of the wallet devoted to the service, instead of a blockchain network, because certain blockchain systems do not support direct queries of this information.\\
When sending or receiving money, information about the transaction provided by the user or shown by the wallet might be altered, causing a security risk and potentially resulting in the user's money being moved to the account of the adversary. If the user's password input screen and the keyboard are observed during the money transfer, the encoded password might be thieved, which violates confidentiality. Diao et al. \cite{Diao}, derived unlock pattern of the user and the status of the foreground program without any authorization, revealing the intensity of security weaknesses in the transmission procedure. If an intruder can disrupt the money transfer or query of the balances or transactions by blocking the link between the wallet and its server or the blockchain network, postures a vulnerability to availability and may result in serious operations like the user extracting the private key to gain back administration of the account, resulting in more impairment. While looking up payments and account balances, an attacker also might deceive the users by falsifying the transmitted information between the wallet server and client, displaying data on the wallet, or data kept on the wallet's server. In this case, the wallet's integrity will be compromised, which results in a display of incorrect transaction registers or incorrect balances on the wallet, consequently deceiving the users \cite{Daojing}. \\

\textbf{Key Management:} If the user has not created a cryptocurrency account, the wallet will randomly produce a couple of private and public keys for a new account on the local device. If the user owns an account, can import the account's private key into the wallet, which enables control of the account from the wallet. Then, the created or imported private key gets encrypted by the digital wallet using user's encryption password. The users might lose full control of their account forever if they lose the private key which leads to the loss of their funds. As a result, the private key should frequently be extracted for backup purposes.\\
If the random seed employed for producing a private key can be anticipated or retrieved during the creation process, the created private key is the potential to be compromised. If the saved private key gets decrypted or retrieved in plaintext during the storage process, it can be stolen and exploited, putting confidentiality at risk. Another way of violating confidentiality occurs when the attacker observes the input of the user and gains the key when the user is manually typing or copying and pasting the key. Moreover, the wallet may show information related to the key on its screen while importing and exporting keys so that an attacker could watch the data in order to achieve the key, endangering confidentiality. Furthermore, when the password for key encryption is configured, an attacker can obtain it by eavesdropping on the user's input, posing a danger to confidentiality. On the other hand, the account's integrity and availability may be at risk if a third party can manipulate or remove the saved key \cite{Daojing}. \\

In the following section, some of the security threats against mobile wallets have been discussed.
\subsection{Digital Wallet's Common Threats}
\textbf{Inappropriate Usage of Platform:} Android and Apple IOS, for example, supply a group of functions of the host operating system. Abusing these services may cause security risks. All of the host system's services have presented implementation rules, and breaking these instructions is the most typical manner of imposing a recognized threat. For instance, using App Local Storage instead of utilizing IOS Keychain to store confidential information in IOS apps. The data stored in the app's local storage may be exposed to other parts of the program, but the data kept in the Keychain is protected from illegal access by the operating system \cite{Sai}.\\

\textbf{Unsafe Data Storage:} Unintended information disclosures and risky data storage fall under this category. If an attacker obtains access to the system, data saved locally in SQL databases and log files may be at risk. External storage of crucial data is recognized as unsafe and can be misused. The detection of unintentional data leaks is not as easy as the detection of intentional leaks. \\Data leaking might be caused by flaws in rooted devices, hardware, or frameworks. Data leakage vulnerabilities can be exploited in applications that lack sufficient monitoring measures of data leaking.\\

\textbf{Inadequate Cryptography:} Cryptographic functions are frequently used in programs that require encryption. Inadequate cryptography can be exploited by two sorts of threats such as weakness in the encryption process and damaged cryptography functionalities. The first is gaining access to confidential information by exploiting a flaw in the construction of the encryption/decryption procedure. The second risk derives from the use of compromised functions of cryptography.\\

\textbf{Reverse engineering:} Like data, reverse engineering targets encryption keys and hardcoded passwords. This approach entails extracting source code from a digital wallet as well as numerous resources from an APK file. These attacks can be accomplished only by hackers who have a deep knowledge of digital wallets \cite{Shukur}. \\

\textbf{Public Wi-Fi:} Using public Wi-Fi such as to conduct digital wallet money transfers can allow third parties to disrupt communication and possibly disrupt payment via MITMF, Wi-Fi sniffing, and DNS spoofing \cite{Kamatchi} \cite{Bosamia}. For instance, an attacker could steal the sensitive information of users who are connected to public Wi-Fi such as in cafes.\\

\textbf{Social engineering:} Instead of breaching or employing practical hacking strategies, social engineering is a technique for gaining control over a computer or information of the users by exploiting human psychology. Attackers might sell the information in black markets or use them to make illegal payments. In addition, they can utilize the obtained information as their identity.\\

\textbf{Phishing attacks:} This kind of attack is one of the most frequent attacks where a phishing link is a type of fraudulent access point that attackers exploit to get critical information and private data from users, such as credit card numbers, financial lottery, or SMS. In phishing attacks, attackers try to acquire login information of the user and personal information, putting digital wallet accounts at risk of theft. For example, the Singapore Police Force (SPF) warned people about the growth of the phishing attack in recent months and it has observed about 1200 cases from December 2021 till January 2022. In most cases, victims were called via messaging applications like WhatsApp. During the conversation, they were asked to provide some private information based on the belief that the caller is from one of the Government agencies \cite{Singapore}.

\section{Conclusion}
A cryptocurrency wallet is a software application or a hardware device that provides users the possibility to execute several transactions. 
% There are two kinds of cryptocurrency wallets, hot and cold. How wallets require internet connection to run the transactions whilst cold wallets work offline and do not need any connection. Multisignature, desktop, mobile, and online wallets belong to hot wallet where paper and hardware wallets are considered as cold wallets. 
Users aiming to buy a digital wallet should recognize their needs and objectives before choosing which type to obtain. Data organization as well as speed, security, and the possibility to execute transactions between two clients are pushing digital wallets into more demand. As these wallets become more popular, the security and safety of the wallets become crucial \cite{Universal}. In this study, we have seen that creating a backup of the private key and also encrypting the digital money using hash functions help diminish privacy and security threats as well as system errors. Employing QR codes as cross-verifying cold wallets is another technique for keeping digital currencies safe. The security of digital wallets has the same objectives as other security systems including availability, integrity, and confidentiality. Moreover, the adversary model for cryptocurrency wallets has been discussed in this study where the adversary or attacker aims to violate the security objectives of the digital wallets. Transaction management and key management as two principal features of crypto wallets provide several functionalities such as sending and collecting the tokens and creating and saving the private key. Exploiting these capabilities by attackers may vulnerabilities to blockchain-based wallets. It's critical to reinforce cryptocurrency wallets with the system's updated security standards, avoid infection of the application supply chain, and mitigate repackaging threats in order to ensure wallet security.

\bibliographystyle{splncs04}
\bibliography{Ref}

\end{document}